# Enabling On-demand Guaranteed QoS for Real Time Video Streaming from Vehicles in 5G Advanced with CAPIF & NEF APIs

Pietro Piscione, Leonardo Lossi
Nextworks
Pisa, Italy

Maziar Nekovee, Chathura Galkandage
6G Lab
University of Sussex
Brighton BN1 9RH, UK

Phil O'Connor, Simon Davies
Honda R&D Europe (U.K.) Ltd
Third Floor, Building 1410,
Arlington Business Park,
Theale, Reading, RG7 4SA, UK

*Abstract*— **This paper presents the design and implementation of a Proof of Concept (PoC) that demonstrates how 5G Advanced Network Functions can be integrated with the Common API Framework (CAPIF) to support enhanced connectivity for automotive applications. The PoC shows the continuous monitoring of the mobile network performance and the on-demand and dynamic adaptation of Quality of Service (QoS) for selected 5G User Equipment (UE) video streaming traffic flows using standard 3GPP Network Exposure Function (NEF) APIs exposed via CAPIF. Moreover, traffic flows are redirected to the edge to improve latency and optimize network resource utilization.**

## I. INTRODUCTION (*Heading 1*)

With the emergence of 5G Advanced and future network architectures, efficient deployment of network functions and bi-directional interactions between vertical applications and network, are crucial for ensuring seamless and QoS-guaranteed service delivery. CAPIF is a standardized framework introduced by 3GPP to unify API exposure across telecommunication networks [1]. CAPIF plays a pivotal role in the interaction of applications with 5G network functions by providing a standardized, secure, and scalable framework for API exposure. This paper summarizes our recent work on a Proof of Concept (PoC) implementing functionalities of CAPIF to enable On-demand Guaranteed QoS of Real Time Video Streaming for automotive applications.

## II. OVERVIEW OF CAPIF

CAPIF provides a common interface for API exposure in telecommunication networks, promoting interoperability and security. It simplifies the interaction between 3rd party vertical applications and network functionalities by offering a standardized approach to service and APIs registration, discovery, and access control. The framework supports multiple deployment models, including centralized and distributed architectures, ensuring flexibility for various network operators.

The CAPIF architecture [2] consists of key components such as:

- **CAPIF Core Functions (CCF):** Handles API exposure and service registration.
- **CAPIF API Provider Functions (CAPF):** Manages network functions and API interactions.
- **CAPIF API Invoker Functions (CAIF):** Enables network function consumers to utilize exposed services.

This modular approach enhances scalability and adaptability to diverse network environments.

## III. SYSTEM ARCHITECTURE AND POC IMPLEMNTATION

### A. System architecture

The system architecture, shown in Figure I, represents an end-to-end emulated 5G network that allows to accommodate a video streaming scenario, with a remote user visualizing the video from the camera installed on board a 5G-connected parked vehicle. The 5G network is based on the Free5GC open-source software [3] with an I-UPF for edge application support and the RAN emulated via UERANSIM. The network integrates NEF capabilities, supporting NEF APIs for Policy Decision and Traffic Quality (PDTQ) Policy Negotiation, Application Server Session with QoS and Event Monitoring, implemented through an extended version of the NEF Emulator in [5] as well as Traffic Influence API implemented in the Free5GC software. All these NEF APIs are exposed via CAPIF, implemented with the ETSI OpenCAPIF open-source software [6]. The video streaming client is based on the VLC tool, which has been extended and complemented with a CAPIF API invoker to consume the NEF APIs.

The components of the architecture are deployed in four VMs in an Openstack virtual environment, representing the access network (RAN), the edge segment (Telco Edge Cloud), the centralized core network (Telco Cloud) and the final user environment (Enhanced VLC), respectively (see Figure I). The security camera of the emulated vehicle is connected to a Raspberry Pi4, with UERANSIM UE software installed and connected to the RAN VM, where the UERANSIM gNB is deployed. In turn, the UERANSIM gNB is connected to the I-UPF deployed in the Edge VM through the N3 network and to the 5G CN control plane in the Telco Cloud VM through the N2 network for the control plane signalling (gNB-to-AMF). The I-UPF is connected to the PSA-UPF through the N9 network, as well as to the local N6 (data) network to enable local edge communications towards the Edge RTMP Server. The ETSI OpenCAPIF and the NEF Emulator are deployed in the Telco Cloud VM.

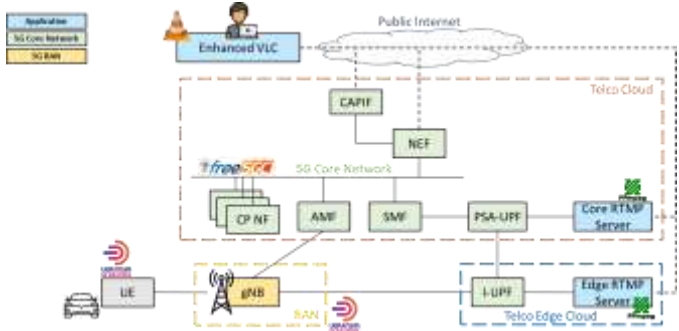

**Figure I System architecture diagram of the PoC.**

*B. PoC implementation and demonstration*

The PoC demonstration starts with an initial workflow for the publication of NEF APIs in CAPIF and their discovery by the API invoker in the enhanced video client application. This is followed by two exemplars scenarios where the application consumes the NEF APIs to guarantee a suitable QoS for the video streaming traffic coming from the 5G connected car. Furthermore a Scenario 1, corresponding to no guaranteed QoS is used as benchmark. The following scenarios have been demonstrated.

Scenario 2: Video streaming with guaranteed QoS on already congested 5G network. The user activates the video streaming when the network is in a congested condition. Interacting with the 5G network via NEF APIs, the video client detects the poor quality of the connectivity to the car UE, and it immediately requests the QoS upgrade for guarantees data rate in uplink, overcoming the initial network congestion.

Scenario 3: Video streaming with dynamic guaranteed QoS upon 5G network congestion. The user activates the video streaming when the network is in an uncongested condition, initially visualizing the video with the desired QoE without the need to request a QoS upgrade. However, to deal with potential future downgrades, the video client uses NEF APIs for QoS monitoring in the target cell and, in case of congestion, requests the QoS upgrade. This scenario demonstrates that, despite the network traffic congestion gradually introduced by other UEs joining the same cell of car UE, the application can dynamically react requesting an adaptation of the network QoS to guarantee a seamless experience to the final user.

## IV. PRELIMINARY RESULTS

The use case scenario assumes a video streaming traffic of a maximum of 4.5 Mbps from the vehicles, and a background traffic generated through Iperf of a maximum of 10 Mbps.

Figure II shows the performance of the video streaming from the vehicle as the network congestion increases in **Scenario 1** (benchmark) and **Scenario 3,** respectively. In Scenario 1 the performance degrades with the increased congestion as the video streaming traffic need to share bandwidth with more users in the cell. In contrast, in Scenario 3 the performance initially degrades with increased network congestion. However, once a user-defined lower threshold in performance is reached, the on-demand QoS is automatically invoked via the NEF APIs, hence bringing back the performance to the pre-defined level. Scenario 3 considers two cases: vehicle located near the cell edge (green curve) and vehicle located near the centre of the cell (blue curve). As expected, performance degradation is more severe when the vehicle is at the cell edge. However, in both cases invoking the QoD process allows to overcome the performance degradation.

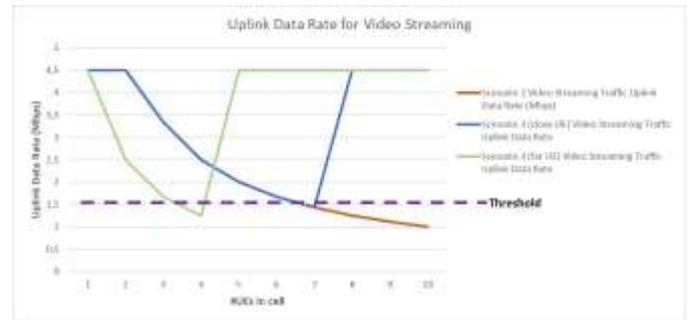

**Figure II Uplink data rate for video streaming in scenarios 1 and 2 is shown as function of network congestion.**

## V. CONCLUSION

This paper describes a software PoC which demonstrates how 5G Advanced Network Exposure can be integrated with the Common API Framework (CAPIF) to support enhanced connectivity for automotive applications. As an example, we demonstrated how QoS can be continuously monitored and enforced on-demand in a video streaming application for an automotive security use case.

## ACKNOWLEDGMENT

This work was jointly funded by Honda R&D Europe (U.K.) Ltd and University of Sussex (through a HEIF Grant).